\documentclass[11pt,a4paper]{article}
\pdfoutput=1

\usepackage{jheppub}
\usepackage[utf8x]{inputenc}
\usepackage[english]{babel}
\usepackage{mathrsfs}
\usepackage{amssymb}
\usepackage{amsmath}
\usepackage{amsthm}
\usepackage{slashed}
\usepackage{pgf}
\usepackage{pdfpages}
\usepackage{times}
\usepackage{color}
\usepackage{hyperref}
\usepackage{emptypage}
\usepackage{epstopdf}
\usepackage{pstool}

\usepackage{epsfig,latexsym,amsfonts,amsmath,amsthm,amssymb,amsbsy,multirow,slashed,color,mathrsfs,wasysym,textcomp,subfigure,wrapfig,comment}

\usepackage{color,datetime,graphicx}

\newcommand{\pink}[1]{\textcolor{\pink}{#1}}

\newcommand{\ddm}{\delta m}

\title{Instabilities of microstate geometries with antibranes}

\preprint{IPhT-T15/185}

\author{Iosif Bena and}
\author{Giulio Pasini}

\affiliation{Institut de Physique Th\'eorique, CEA Saclay, CNRS UMR 3681,
 91191 Gif-sur-Yvette, France}

 \emailAdd{iosif.bena@cea.fr, giulio.pasini@cea.fr}

\abstract{One can obtain very large classes of horizonless microstate geometries corresponding to near-extremal  black holes by placing probe supertubes whose action  has metastable minima inside certain supersymmetric bubbling solutions~\cite{Bena:supertube_microstate}. We show that these minima can lower their energy when the bubbles move in certain directions in the moduli space, which implies that  these near-extremal microstates are in fact unstable once one considers the dynamics of all their degrees of freedom. The decay of these solutions corresponds to Hawking radiation, and we compare the emission rate and frequency to those of the corresponding black hole. Our analysis supports the expectation that generic non-extremal black holes microstate geometries should be unstable. It also establishes the existence of a new type of instabilities for antibranes in highly-warped regions with charge dissolved in fluxes. 
}
 
\begin{document}
\maketitle

\section{Introduction}

Recent information-theory arguments (that can be grouped under the fuzzball or firewall labelings) indicate that the classical description of black holes breaks down at horizon scale and has to be replaced by a horizonless structure that allows Hawking radiation to carry away the information from the black hole~\cite{Mathur:information, Mathur:2009hf, Braunstein:2009my, AMPS, AMPS2}.

The most advanced effort to build this kind of horizonless structure  in string theory, oftentimes known as the ``microstate geometry'' programme, consists of constructing smooth supergravity solutions that have the same charge, mass and angular momenta of the black hole but have no event horizon\footnote{For other recent attempts to construct stringy structure at the black hole horizon see \cite{Dodelson:2015toa, PRU}.}. In a microstate geometry the would-be black hole throat is capped and instead of a horizon one has a number of nontrivial cycles (bubbles) threaded by fluxes
\cite{Giusto:2004id,Giusto:2004ip,Bena_bubbling_supertubes, Berglund, Bena_foaming, Bena:mergers, Bena:abyss}. Thus, the singularity of the black hole would be resolved at the scale of the horizon by a geometric transition analogous to the Klebanov-Strassler~\cite{Klebanov}, Polchinski-Strassler~\cite{Polchinski-Strassler} and Lin-Lunin-Maldacena~\cite{LLM} ones.

An elaborate technology has been developed to construct smooth microstate solution for supersymmetric three- and four-charge black holes in the framework of 11D or type II supergravities compactified to five or six-dimensions, and by now very large families of such solutions are known\footnote{The largest family, parameterized by arbitrary continuous functions of two variables, was recently constructed in \cite{Bena:superstrata}.}. Furthermore, the entropy of these solutions has been argued to reproduce the growth with charges of the Beckenstein-Hawking entropy of the corresponding black hole~\cite{superstrata-counting}, so there is little doubt left that the horizons of supersymmetric black holes will be replaced by horizonless structure.

However, despite this success, the world of non-BPS non-extremal microstates remains quite unexplored. The few known exact solutions in the JMaRT class~\cite{jmart, giusto_ross_saxena, AlAlawi:2009qe, Banerjee:2014hza, bossard-katmadas-recent} and the Running-Bolt class~\cite{running_bolt, bossard-katmadas-bolt}, though horizonless, do not have the right charges to correspond to a black hole with a classically-large horizon area. Furthermore, their construction is quite artisanal at this point, although the recent discovery of the Bossard-Katmadas (or the ``Floating JMaRT'') system \cite{bossard-katmadas-recent} indicates that more systematic methods may be on the horizon \cite{bossard-katmadas-next}.

In parallel to these efforts, it was argued in~\cite{Bena:supertube_microstate} that one can systematically obtain very large classes of microstate geometries for non-extremal black holes by placing a supertube~\cite{Mateos} at metastable minimum inside the known BPS microstate geometries. The energy $\Delta M$ of this supertube then gives the mass above extremality of the solution. Although no exact solutions in this class are known, one can argue that the supertubes should backreact into a smooth solution\footnote{For supersymmetric solutions it was shown that the Born-Infeld equations governing the supertube are equivalent to those ensuring smoothness and absence of closed timelike curves in the fully-backreacted solution \cite{Ruef-DBI-meets-sugra}.}, and hence one expects that there should exist a very large number of bubbled microstate geometries corresponding to near-extremal black holes. Since supertubes have charges opposite to those of the black hole, these bubbled geometries would have cycles that are wrapped both by positive and by negative fluxes\footnote{Similar to the proposed metastable flux compactifications of \cite{aganagic_beem}}, and this is exactly the structure that Gibbons and Warner have proven to be necessary if one is to replace non-extremal black holes by stationary horizonless black hole solitons \cite{Gibbons:2013tqa,Haas:2014spa,deLange:2015gca}. 

Furthermore, since these microstate geometries have neither inner nor outer horizon, they should be thought of as resolving the singularity of the non-extremal black hole all the way to the outer horizon, which is {\em backwards in time} from the location of the singularity. This pattern of singularity resolution is rather extraordinary and, if confirmed by the construction of fully backreacted non-extremal microstate geometries, it would have important implications not only for black hole singularities but also for cosmological ones. 

There is another important difference between the non-extremal microstate geometries in the JMaRT and Running Bolt classes, and the near-extremal microstates constructed using anti-supertubes. The JMaRT geometry is unstable \cite{Dias-Hovdebo-Myers} (and so is the Runnning Bolt one \cite{avery_unpublished}) and this instability, which comes from the existence of an ergoregion, can be matched precisely to the fact that the D1-D5-P CFT state dual to the JMaRT geometry is also unstable \cite{Chowdhury:2007jx, Avery-Chowdhury-Mathur}. Furthermore, it has been argued that a similar ergoregion instability should be present in all non-extremal microstate geometries \cite{Chowdhury-Mathur}.

On the other hand, the near-extremal microstate geometries of~\cite{Bena:supertube_microstate} are obtained by placing probe anti-supertubes  inside extremal bubbling geometries at metastable minima of their Hamiltonian, and hence these configurations could in principle be much longer lived than one expects for a typical near-extremal microstate by studying the D1-D5 CFT. Indeed, the near-extremal CFT states consist of a very large number of (supersymmetric) left-mover momentum modes and a much smaller number of supersymmetry-breaking right movers, and it seems very difficult to prepare states where the annihilation of these modes is suppressed such that the decay takes place over very long time scales. 

In this paper we resolve this tension by showing that the near-BPS microstate geometries that one obtains by placing metastable anti-supertubes inside long scaling solutions~\cite{Bena:supertube_microstate} can in fact lower their energy when the bubbles of the scaling solution move relative to each other. Hence, what appears to be a metastable configuration from the point of view of the action of a probe brane is in fact an unstable one if one takes into account the degrees of freedom corresponding to the motion of the bubbles. 

We study a scaling microstate geometry that is constructed using seven collinear Gibbons-Hawking centers \cite{Bena:mergers} as well as an anti-supertube probe. If one keeps the GH centers collinear, the supertube Hamiltonian will have both supersymmetric and metastable minima~\cite{Bena:supertube_hamiltonian}, corresponding respectively to microstate geometries for supersymmetric and non-extremal black holes. For the latter, the mass above extremality of the black hole, $\Delta M$, is simply equal to the value of the Hamiltonian in the metastable minimum. 

However, it is well known that supersymmetric solutions that are constructed using $N$ GH centers have a $2N-2$-dimensional moduli space, that is parameterized by the solutions of the $N-1$ bubble (or integrability) equations that govern the inter-center distances \cite{bates_denef, Bena_bubbling_supertubes, Berglund}. Hence, the mass above extremality, $\Delta M$, is in fact a function of the position in the moduli space. Thus, if one is to prove that the supertube minimum corresponds to a metastable black hole microstate, one must also show that $\Delta M$ has a minimum when the GH points are collinear, and does not decrease as one moves around the moduli space\footnote{Things are even a bit more complicated: as the GH points move in the $2N-2$-dimensional space of solutions to the bubble equations the $SU(2)$ angular momentum of the BPS solution, $J_L$, changes and hence the solutions do not correspond to the same black hole. For fixed values of the charges and angular momenta the moduli space of microstates of the corresponding black hole is a constant-$J_L$ slice of the $2N-2$ space of solutions to the bubble equations, and has therefore dimension $2N-5$.}. We show that it does.

Indeed, if one examines $\Delta M$ as a function of certain moduli space directions one finds that the collinear GH configuration corresponds to a saddle point, and $\Delta M$ can in fact decrease as some of the centers move off the axis. Our result implies that the instability is triggered by the motion of the bubbles threaded by flux. As the configuration moves away from the saddle point, the kinetic energy of the bubbles increases, and we expect this to result in gravitational radiation that will relax the system towards a supersymmetric minimum.

One can also estimate the time-scale characteristic to this instability, as well as the energy emission rate, as a function of the charges and mass of the solution as well as of the other parameters of the non-extremal microstate geometry.  To compute the time-scale we estimate the energy of the configuration as a function of the tachyonic rotation angle in the GH base space of the solution, as well as the kinetic term corresponding to the motion in the moduli space. The latter calculation is performed by formally interpreting our solution as a multi-center four-dimensional solution whose constant in the Taub-NUT (D6 brane) harmonic function has been set to zero. We then argue that the kinetic term corresponding to the angular motion we consider is the same as the one obtained by replacing some of the centers of the bubbling solution with the corresponding black hole and black ring. This allows us to compute this term and to find its scaling with the length of the black hole microstate throat.

We also check whether the microstate decay channel we study is similar to the decay channel one expects for a typical microstate of a D1-D5-P near-extremal black hole, which was computed in~\cite{callan} and nicely follows from Stephen's law in five dimensions. There are three quantities that one can compare: the emission rate, the frequency peak energy and the typical radius of the microstate~\cite{Bena:abyss}. By changing the parameters of the microstate solutions we scan over, we can get any two of these parameters to agree, but not the third. This indicates that the particular non-extremal microstate geometries that we are considering are not typical. 

This is not at all unexpected: of all the microstates of the BPS black hole we have chosen to uplift to a non-BPS one only a particular one, corresponding to seven collinear GH centers with certain fluxes on them. Furthermore, the starting configuration is clearly not a typical representative of the microstate geometries of the BPS black hole - one can find even more complicated solutions with a GH base where the centers are not aligned, and we expect from \cite{superstrata-counting} that the typical states that contribute to the entropy of this black hole come from superstrata excitations of solutions with a GH base, that depend on arbitrary functions of two variables. Hence, it is hard to expect generically that the uplift of the non-typical seven-center BPS microstate geometry will give us a typical non-BPS microstate geometry whose decay rate will match that of the black hole. 
 
Our calculations does open the way to investigate whether in the phase space of bubbling solutions we can find non-BPS microstate geometries that decay as the typical microstates do. There are two possibilities: either the typical microstates have many more bubbles than the solution we consider (and then our calculations should reveal that the decay rate approaches that of the black hole as one increases the number of bubbles), or the typical microstates will have a smaller number of bubbles, and their entropy will come from the oscillations of these bubbles that leave the topology intact. The recent superstratum counting of \cite{superstrata-counting} appear to favor the second possibility, although there are quite a few subtleties that may incline the balance in favor of the first one.

Arguments about typicality aside, we believe that our investigation has two important conclusions: The first is that microstate geometries of non-extremal black holes will be unstable, and therefore the dynamics of these black holes will correspond to a chaotic motion of mutually-non-supersymmetric centers at a bottom of a black-hole like throat. It would be interesting to investigate whether one can obtain generic features of this chaotic behavior, and perhaps compare it to other situations where chaotic behaviors appear in black hole physics \cite{Shenker}.

The second implication of our calculations is the discovery of a new decay channel for antibranes in solutions with charge dissolved in fluxes. Normally one studies these antibranes by considering them as probes in a solution and examining their action while assuming that the solution remains unchanged. Our investigation shows that this approach can give misleading results, and in order to determine whether an antibrane is metastable or unstable one should examine its full interactions with the moduli of the underlying solution. 

This paper is organized as follows. In Section~\ref{section_review} we briefly review the construction of the near-BPS microstates found in~\cite{Bena:supertube_microstate}. In Section~\ref{section_instability} we explicitly show that these near-BPS microstates are unstable along a direction in the moduli space. Then in Section~\ref{section_strategy} we compute the energy emission rate for a decay along the channel corresponding to this direction. In Section~\ref{section_typicality} we study how the emission rate and other quantities scale with the length of the throat and the charges of the solution. We also compare the microstate emission rate, emission frequency and radius to those corresponding to the typical states in the black hole thermodynamic ensemble. We conclude in Section~\ref{section_conclusions} and present some future directions. The detailed construction of a three-charge smooth microstate can be found in Appendix~\ref{appendix_background}, and some of the intuition used to determine the kinetic term corresponding to the motion in the moduli space of BPS solutions is explained in more detail in Appendix~\ref{appendix_physics}.

 \section{Near-extremal black hole microstate geometries}\label{section_review}
 
In this section we briefly review the construction of near-extremal black hole microstate configurations by placing a metastable supertube inside a supersymmetric bubbling solution~\cite{Bena:supertube_microstate}. We first present the main features of a smooth scaling microstate solution that has the charges of a five-dimensional BPS black hole with a classically large horizon area~\cite{Bena:mergers}. We then make the microstate near-extremal by placing an anti-supertube probe at a metastable minimum.\\
 The full features of the BPS three-charge black hole microstate solution are relegated to Appendix~\ref{appendix_background}.

\subsection{Microstate geometries for BPS black holes with large horizon area}\label{initialsetup}
\begin{figure}[h!]
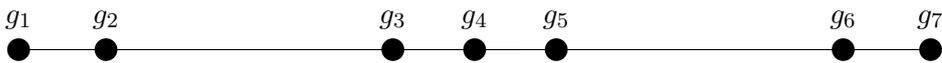

\begin{pgfpicture}{0cm}{0cm}{10cm}{2cm}
\pgfsetxvec{\pgfpoint{0.5cm}{0cm}}
\pgfsetyvec{\pgfpoint{0cm}{0.5cm}}
\pgftranslateto{\pgfxy(14.5,1)}


\pgfcircle[fill]{\pgfxy(-12,0)}{0.15cm}
\pgfputat{\pgfxy(-12,0.8)}{\pgfbox[center,center]{$g_1$}}
\pgfline{\pgfxy(-11.85,0)}{\pgfxy(-9.85,0)}

\pgfcircle[fill]{\pgfxy(-9.7,0)}{0.15cm}
\pgfputat{\pgfxy(-9.7,0.8)}{\pgfbox[center,center]{$g_2$}}
\pgfline{\pgfxy(-9.55,0)}{\pgfxy(-2.3,0)}

\pgfcircle[fill]{\pgfxy(-2.15,0)}{0.15cm}
\pgfputat{\pgfxy(-2.15,0.8)}{\pgfbox[center,center]{$g_3$}}
\pgfline{\pgfxy(-2,0)}{\pgfxy(0.15,0)}

\pgfcircle[fill]{\pgfxy(0,0)}{0.15cm}
\pgfputat{\pgfxy(0,0.8)}{\pgfbox[center,center]{$g_4$}}
\pgfline{\pgfxy(1.25,0)}{\pgfxy(0.15,0)}

\pgfcircle[fill]{\pgfxy(2.15,0)}{0.15cm}
\pgfputat{\pgfxy(2.15,0.8)}{\pgfbox[center,center]{$g_5$}}
\pgfline{\pgfxy(2,0)}{\pgfxy(0.15,0)}

\pgfcircle[fill]{\pgfxy(9.7,0)}{0.15cm}
\pgfputat{\pgfxy(9.7,0.8)}{\pgfbox[center,center]{$g_6$}}
\pgfline{\pgfxy(9.55,0)}{\pgfxy(2.3,0)}

\pgfcircle[fill]{\pgfxy(12,0)}{0.15cm}
\pgfputat{\pgfxy(12,0.8)}{\pgfbox[center,center]{$g_7$}}
\pgfline{\pgfxy(11.85,0)}{\pgfxy(9.85,0)}

\end{pgfpicture}
\caption{The collinear configuration of GH centers that we start from. Note that the distances between the satellites and the central blob are not on scale}\label{figurecolinear}
\end{figure}

A smooth horizonless microstate geometry that has the same charges and supersymmetries of a five-dimensional three-charge BPS black hole and has a Gibbons-Hawking (GH) base space has the metric:
\begin{equation}\label{metric_1}
 ds^2_{11}=-(Z_1Z_2Z_3)^{-\frac{2}{3}}(dt+k)^2+(Z_1Z_2Z_3)^{\frac{1}{3}}ds_4^2+(Z_1Z_2Z_3)^{\frac{1}{3}} 
\sum_{I=1}^{3}\frac{dx_{3+2I}^2+dx_{4+2I}^2}{Z_I}
\end{equation}
where the $Z_I$ are the warp factors corresponding to the three charges, $k$ is the angular-momentum 1-form and $ds^2_4$ is the metric of the GH base:
\begin{equation}\label{gh_1}
ds^2_4=V^{-1} (d\psi+A)^2+V(dy_1^2+dy_2^2+dy_3^2)
\end{equation}
All the quantities appearing in~\eqref{metric_1} and~\eqref{gh_1} are defined following the standard procedure \cite{Bena_foaming,Berglund} detailed in Appendix~\ref{appendix_background}. The whole background is determined once one fixes the number $N$ of GH centers, as well as the residues of the four harmonic functions $V$ and $K^I$ at these centers:
\begin{align}
V=\sum_{i=1}^N \frac{v_i}{r_i} \quad \quad \quad K^I=\sum_{i=1}^N \frac{k_i^I}{r_i}\quad \quad \quad r_i=|\vec{y}-\vec{g}_i|\label{harmonic_functions}
\end{align}
where $\vec{y}=(y_1,y_2,y_3)$ and $\vec{g}_i$ is the position of the i-th pole in the same subspace.
The particular BPS microstate geometry we will use in this paper has $N=7$ GH centers, whose $(v_i,k_i^I)$ parameters are symmetric with respect to the GH center in the middle\footnote{The value of $k_1^3$ given in~\cite{Bena:supertube_microstate} differs from the one we give here and in~\cite{Bena:abyss} because of a typo.}:
\begin{align}
  v_1&=20 \quad \quad v_2=-20 \quad \quad v_3=-12 \quad \quad v_4=25\nonumber \\
k_i^{1}&=\frac{5}{2}|v_i| \quad \quad k_i^2=\hat{k}|v_i| \quad \quad k_i^3 = \frac{1}{3}|v_i| \quad \quad  i=3,4 \nonumber \\
k_1^{1}&=1375 \quad \quad k_1^2=-1835/2+980 \hat{k} \quad  \quad k_1^3 = -8260/3  \nonumber \\
k_2^{1}&=1325 \quad \quad k_2^2=-1965/2-980 \hat{k} \quad \quad  k_2^3 = 8380/3 \nonumber \\
\quad v_{8-i}&=v_i \quad \quad   k^I_{8-i}=k^I_i, \quad \quad i=1,2,3 
 \label{charges}
\end{align}
where the meaning of $\hat{k}$ will become clear in a moment. \\
As explained in Appendix~\ref{appendix_background}, the distances $r_{ij}$ between the GH centers are subject to $N-1$ \textit{bubble equations}~ that need to be satisfied to prevent the existence of closed timelike curves:
\begin{align}\label{bubbleequations}
  \sum_{j=1\, j\neq i}^{N} \Pi_{ij}^{(1)}\Pi_{ij}^{(2)}\Pi_{ij}^{(3)}\frac{v_iv_j}{r_{ij}}=-v_i\sum_{I=1}^3\sum_{s=1}^N k_s^I+\sum_{I=1}^{3}k_i^I  \quad \quad \quad {\rm with}\quad \quad \Pi_{ij}^I\equiv\left(\frac{k^I_j}{v_j}-\frac{k^I_i}{v_i} \right)
  \end{align}
To solve~\eqref{bubbleequations} with the parameters~\eqref{charges} we first constrain all the GH centers to lie on the same axis. Given that the parameters in~\eqref{charges} are invariant under $i\rightarrow 8-i$ for $i=1,...4$, the solution of~\eqref{bubbleequations} will give rise to a collinear configuration, shown in Figure~\ref{figurecolinear}, with
\begin{equation}\label{z2symmetry}
r_{ij}=r_{(8-i)(8-j)}
\end{equation}
Hence, the collinear solution is determined completely by $r_{12},r_{23},r_{34}$.\\

The family of collinear microstate geometries we consider is parameterized by $\hat{k}$, which controls the depth of our microstate. There is a critical value $\hat{k}_0$ at which the solution is singular\footnote{For the curious, $\hat{k}_0\approx 3.17975$.}, and changing $\hat{k}$ around this value gives rise to scaling solutions that have very small $r_{ij}$ and hence a very long throat:
\begin{align}\label{scaling}
 \hat{k} = \hat{k}_0+ \epsilon\,, \quad \quad \quad r_{ij}=\epsilon \overline{r}_{ij} + \mathcal{O}(\epsilon^2)\,,
\end{align}
where the $\overline{r}_{ij}$ are determined by the fluxes on the two-cycles between the centers. As expected, in the scaling limit the ratios between the distances are fixed \cite{bates_denef}, and the physical distances between the GH centers become independent of the distance between these points in the GH base \cite{Bena:mergers}. However, as $\epsilon$ approaches zero the length of the throat of the microstate geometry diverges as $\epsilon^{-1}$. 

We choose to work with a throat that is long-enough to describe the typical sector of the D1-D5-P black hole but not infinite, and we will fix the length of the throat for now by setting $\hat{k}=3.1667$. In Section \ref{section_typicality} we will relax this condition and examine how the physics we find changes as $\hat{k}$ moves towards the critical value. We then use~\eqref{charges} to solve~\eqref{bubbleequations} and find
\begin{align}
r_{12}=3.58 \, \cdotp 10^{-3} \quad r_{23}=23.84, \quad r_{34}=5.78 \, \cdotp 10^{-3} 
\end{align}
and the ratios
\begin{align}
 \frac{r_{23}}{r_{12}}\sim 6.7 \, \cdotp 10^3 \quad  \frac{r_{23}}{r_{34}}\sim 4.1\, \cdotp 10^3 \label{ratios}
 \end{align}
We discuss in Appendix~\ref{appendix_physics} the physical interpretation of these very large ratios. 

The three electric charges of the solution and its $SU(2)$ angular momenta are~\eqref{bhchargesfixed}:
\begin{align}
Q_1=1.48 \, \cdotp 10^5, \quad Q_2=1.20 \, \cdotp 10^5, \quad Q_3= 1.76\, \cdotp 10^5, \quad J_R=1.018 \, \cdotp 10^8, \quad J_L=0. \label{chargesmacroscopic}
\end{align}
As one can see from equation~\eqref{Jl},  $J_L$ vanishes because of the $Z_2$ symmetry of our configuration~\eqref{z2symmetry}. 
This solution represents a supersymmetric horizonless microstate of a BMPV black hole~\cite{BMPV} with a classically large horizon area (and hence nonzero entropy)~\cite{Bena:mergers}.

\subsection{Adding Metastable Supertubes}\label{section_supertube}

To build a near-BPS microstate we add a supertube probe~\cite{Mateos} to the BPS solution of Section~\ref{initialsetup}. If the supertube is at a supersymmetric minimum, the resulting microstate is still BPS. However, supertubes can also have metastable minima~\cite{Bena:supertube_hamiltonian}, and these give rise to microstate geometries of near-BPS black holes~\cite{Bena:supertube_microstate}.\\
In the duality frame where the black hole has three M2 brane charges~\eqref{metric_1}, a supertube has two types of M2 brane charges, $q_1$, $q_2$, as well as a dipole charge $d_3$ corresponding to an M5 brane that wraps the fiber of the Gibbons-Hawking space~\cite{Bena_bubbling_supertubes}. The potential energy of a supertube is~\cite{Bena:supertube_hamiltonian}:
\begin{equation}\label{hamiltonian}
\mathcal{H}=\frac{\sqrt{Z_1Z_2Z_3V^{-1}}}{d_3\rho^2}\sqrt{\left(\tilde{q}_1^2+d_3^2\frac{\rho^2}{Z_2^2}\right)\left(\tilde{q}_2^2+d_3^2\frac{\rho^2}{Z_1^2}\right)}+\frac{\mu \tilde{q}_1\tilde{q}_2}{d_3\rho^2}-\frac{\tilde{q}_1}{Z_1}-\frac{\tilde{q}_2}{Z_2}-\frac{d_3 \mu}{Z_1Z_2}+q_1+q_2
\end{equation}
where
\begin{align}
\tilde{q}_1\equiv q_1+d_3(K^2 V^{-1}-\mu/Z_2), \quad \tilde{q}_2\equiv q_2+d_3(K^1 V^{-1}-\mu/Z_1)
\end{align}
and $\rho$ is proportional to the size of the GH fiber at the location of the supertube
\begin{equation}
\rho^2 \equiv Z_1Z_2Z_3V^{-1}-\mu^2
\end{equation}
The warp factors $Z_I$ and the angular momentum parameter $\mu$ are defined in~\eqref{warpfactors} and~\eqref{mu} respectively, and $V$ and $K^I$ are the harmonic functions defined in~\eqref{harmonic_functions}. When the supertube is at a metastable point of the potential, $\mathcal{H}>q_1+q_2$, and supersymmetry is broken. The total charges of the system are then given by the sum of the charges of the background and of the probe, while the total mass is
  \begin{equation}\label{mass_almost_bps}
  M_{tot}=\Delta M + \sum{Q_{background}+q_{probe}}
\end{equation}
where $\Delta M$ is the value of $\mathcal{H}-(q_1+q_2)$ at the metastable point. Even if the backreacted solution corresponding to this supertube has not been constructed explicitly, it is possible to argue that the resulting background is globally smooth in the duality frame where the charges of the black hole correspond to D1 branes, D5 branes and momentum. 

As in ~\cite{Bena:supertube_microstate}, we consider a supertube whose charges are much smaller than those of the background, and whose physics can  therefore be captured by the probe approximation:
\begin{equation}\label{chargesupertube}
(q_1,q_2,d_3)=(10,-50,1)
\end{equation}

The fact that $q_2$ and $q_1$ have opposite signs does not automatically imply that supersymmetry is broken. A supertube with a given set of charges can have both BPS and metastable minima, and the parameters whose positivity ensures that supersymmetry is not broken are the $\tilde q_i$.

 Let $y_1$ be the coordinate parameterizing the axis of Figure~\ref{figurecolinear} centered in $g_4$. Given the symmetric arrangement of the GH points in Section~\ref{initialsetup}, $\mathcal{H}$ in~\eqref{hamiltonian} is invariant under $y_1 \rightarrow -y_1$. The probe potential has several Mexican-hat-type metastable minima, and we focus on the one in the proximity of $g_6$. The mass above extremality of this supertube is
\begin{equation}\label{metastable0value}
 \Delta M = \mathcal{H}(23.812,0,0)-(q_1+q_2)\sim 0.04742
 \end{equation}
 
It is important to stress that the existence of metastability does not depend on the parameter $\epsilon$ controlling the depth of the scaling BPS microstate geometry, because when the depth scales with $\epsilon$ as in~\eqref{scaling} the probe potential~\eqref{hamiltonian} transforms as 
   \begin{equation}\label{hamiltonian_scaling}
   \mathcal{H}( r_{ij}^\prime)\rightarrow \epsilon \mathcal{H}(r_{ij})+\mathcal{O}(\epsilon^2)
   \end{equation}


\section{The instability of near-extremal microstates}\label{section_instability}

In this section we show that the nearly-BPS microstates built in Section~\ref{section_supertube} by using probe anti-supertubes are classically unstable.

\begin{figure}[h!]
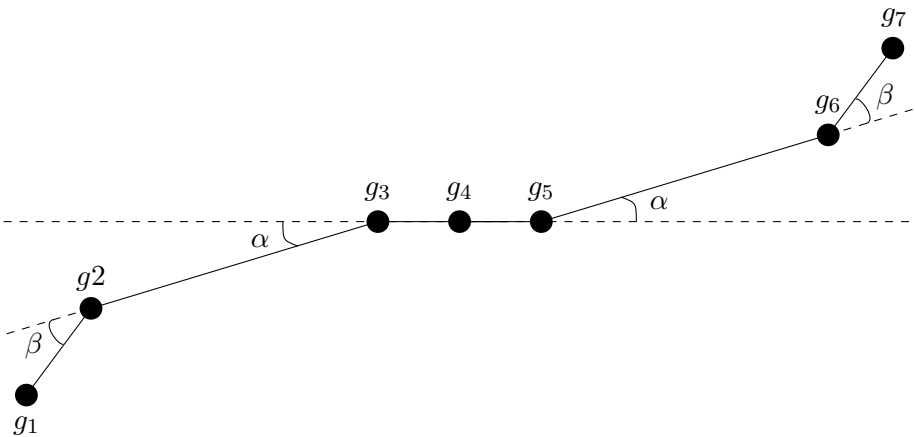

\begin{pgfpicture}{0cm}{0cm}{10cm}{5.5cm}
\pgfsetxvec{\pgfpoint{0.5cm}{0cm}}
\pgfsetyvec{\pgfpoint{0cm}{0.5cm}}
\pgftranslateto{\pgfxy(15,5)}

\pgfcircle[fill]{\pgfxy(-11.4,-4.6)}{0.15cm}
\pgfputat{\pgfxy(-11.4,-5.4)}{\pgfbox[center,center]{$g_1$}}

\pgfputat{\pgfxy(-9.7,-2.3)}{

\pgfmoveto{\pgfxy(-1,-0.3)}
\pgfcurveto{\pgfxy(-1.3,-0.4)}{\pgfxy(-0.85,-1)}{\pgfxy(-0.73,-0.95)}
\pgfstroke
\pgfputat{\pgfxy(-1.5,-1)}{\pgfbox[center,center]{$\beta$}}

}

\pgfline{\pgfxy(-9.7,-2.3)}{\pgfxy(-11.4,-4.6)}
\pgfline{\pgfxy(-9.7,-2.3)}{\pgfxy(-2.15,0)}

\pgfcircle[fill]{\pgfxy(-9.7,-2.3)}{0.15cm}
\pgfputat{\pgfxy(-9.7,-1.5)}{\pgfbox[center,center]{$g2$}}

\pgfcircle[fill]{\pgfxy(-2.15,0)}{0.15cm}
\pgfputat{\pgfxy(-2.15,0.8)}{\pgfbox[center,center]{$g_3$}}
\pgfline{\pgfxy(-2,0)}{\pgfxy(0.15,0)}

\pgfcircle[fill]{\pgfxy(0,0)}{0.15cm}
\pgfputat{\pgfxy(0,0.8)}{\pgfbox[center,center]{$g_4$}}
\pgfline{\pgfxy(1.25,0)}{\pgfxy(0.15,0)}

\pgfcircle[fill]{\pgfxy(2.15,0)}{0.15cm}
\pgfputat{\pgfxy(2.15,0.8)}{\pgfbox[center,center]{$g_5$}}
\pgfline{\pgfxy(2,0)}{\pgfxy(0.15,0)}

\pgfcircle[fill]{\pgfxy(9.7,2.3)}{0.15cm}
\pgfputat{\pgfxy(9.7,3.1)}{\pgfbox[center,center]{$g_6$}}

\pgfcircle[fill]{\pgfxy(11.4,4.6)}{0.15cm}
\pgfputat{\pgfxy(11.4
4,5.4)}{\pgfbox[center,center]{$g_7$}}

\pgfputat{\pgfxy(9.7,2.3)}{

\pgfmoveto{\pgfxy(1,0.3)}
\pgfcurveto{\pgfxy(1.3,0.4)}{\pgfxy(0.85,1)}{\pgfxy(0.73,0.95)}
\pgfstroke
\pgfputat{\pgfxy(1.5,1)}{\pgfbox[center,center]{$\beta$}}

}

\pgfline{\pgfxy(9.7,2.3)}{\pgfxy(11.4,4.6)}
\pgfline{\pgfxy(9.7,2.3)}{\pgfxy(2.15,0)}

\pgfputat{\pgfxy(-2.15,0)}{

\pgfmoveto{\pgfxy(-2.5,0)}
\pgfcurveto{\pgfxy(-2.5,-0.5)}{\pgfxy(-2.5,-0.5)}{\pgfxy(-2.1,-0.65)}
\pgfstroke
\pgfputat{\pgfxy(-3.1,-0.5)}{\pgfbox[center,center]{$\alpha$}}

}

\pgfputat{\pgfxy(2.15,0)}{

\pgfmoveto{\pgfxy(2.5,0)}
\pgfcurveto{\pgfxy(2.5,0.5)}{\pgfxy(2.5,0.5)}{\pgfxy(2.1,0.65)}
\pgfstroke
\pgfputat{\pgfxy(3.1,0.5)}{\pgfbox[center,center]{$\alpha$}}

}

\pgfsetdash{{3pt}{3pt}}{0pt}
\pgfline{\pgfxy(-12,0)}{\pgfxy(12,0)}
\pgfline{\pgfxy(9.7,2.3)}{\pgfxy(12,3)}
\pgfline{\pgfxy(-9.7,-2.3)}{\pgfxy(-12,-3)}

\end{pgfpicture}
\caption{The parameterization for planar symmetric configurations with seven centers} \label{parametrization}
\end{figure}

We consider a microstate geometry with seven GH centers that have the $v_i$ and $k_i^I$ parameters as in~\eqref{charges}. The location of the GH centers is constrained by the bubble equations~\eqref{bubbleequations} and by the requirement that $J_L$ is zero, and hence the moduli space of the solutions is six-dimensional. For simplicity we examine a subset of this moduli space constrained by the symmetry~\eqref{z2symmetry}, which automatically ensures that $J_L=0$ and reduces the number of independent bubble equations from six to three. Furthermore, we will focus our discussion only on configurations where all the centers lie on the same plane\footnote{We have also analyzed the configurations where the centers are not on the same plane, but all the relevant physics is captured by the planar ones.}. The multicenter solutions that satisfy all these requirements can be parameterized by two coordinates $(\alpha, \beta)$, that are the angles shown in Figure~\ref{parametrization}. 

As one changes these angles $(\alpha, \beta) \in [0,2\pi]\times[0,\pi]$, the bubble equations~\eqref{bubbleequations} determining the inter-center distances are modified, and have to be solved again to determine the new values of the distances. In principle this way of parameterizing the solutions of the bubble equations can lead to singularities, as the bubbles can collapse for certain critical values of the angles, but this does not happen for the particular solution we consider.

It is not hard to see first that as one changes $\alpha$ and $\beta$ the action of a probe supertube with charges~\eqref{chargesupertube} will continue having a metastable minimum in the vicinity of $g_6$ (or $g_2$). However, the exact value of the energy of the probe, which gives the mass above extremality, becomes a nontrivial function of the angles:  $\Delta M(\alpha, \beta)$.

\begin{figure}[h!]
\centering
\includegraphics[width=0.8\textwidth]{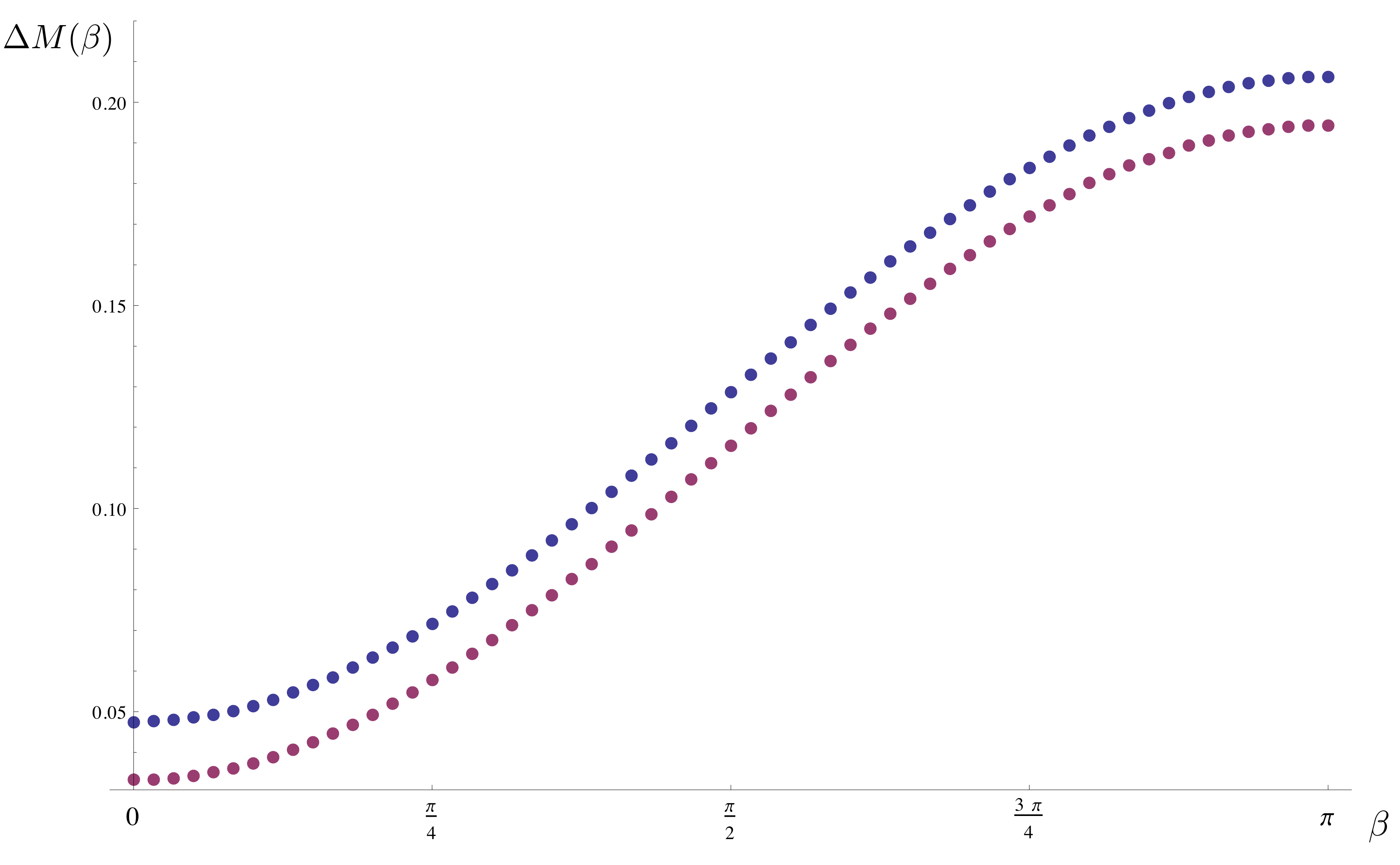}
 \caption{Plots of 60 values for $\Delta M$ in the interval $\beta \in [0,\pi]$ for $\alpha=0$ in two different scaling regimes. The blue curve is obtained for $\hat{k}=3.1667$, while the purple one is obtained for $\hat{k}=3.175$, which corresponds to $\epsilon = 0.36$ in~\eqref{scaling}. For the sake of clarity, the values of the purple curves have been multiplied by $0.98$ times the ratio of the two $\Delta M$ for the collinear configurations.}
 \label{beta_trend}
 \end{figure}

We analyzed $\Delta M(\alpha, \beta)$ numerically starting from the collinear configuration $(\alpha,\beta)=0$. We found that keeping $\alpha$ fixed $\Delta M$ is monotonically increasing with $\beta\in[0,\pi]$ and this behavior does not depend on the choice of $\hat{k}$ for the scaling~\eqref{scaling}, as shown in Figure~\ref{beta_trend}. 
\begin{figure}[h!]
\centering 
\includegraphics[width=.8\textwidth]{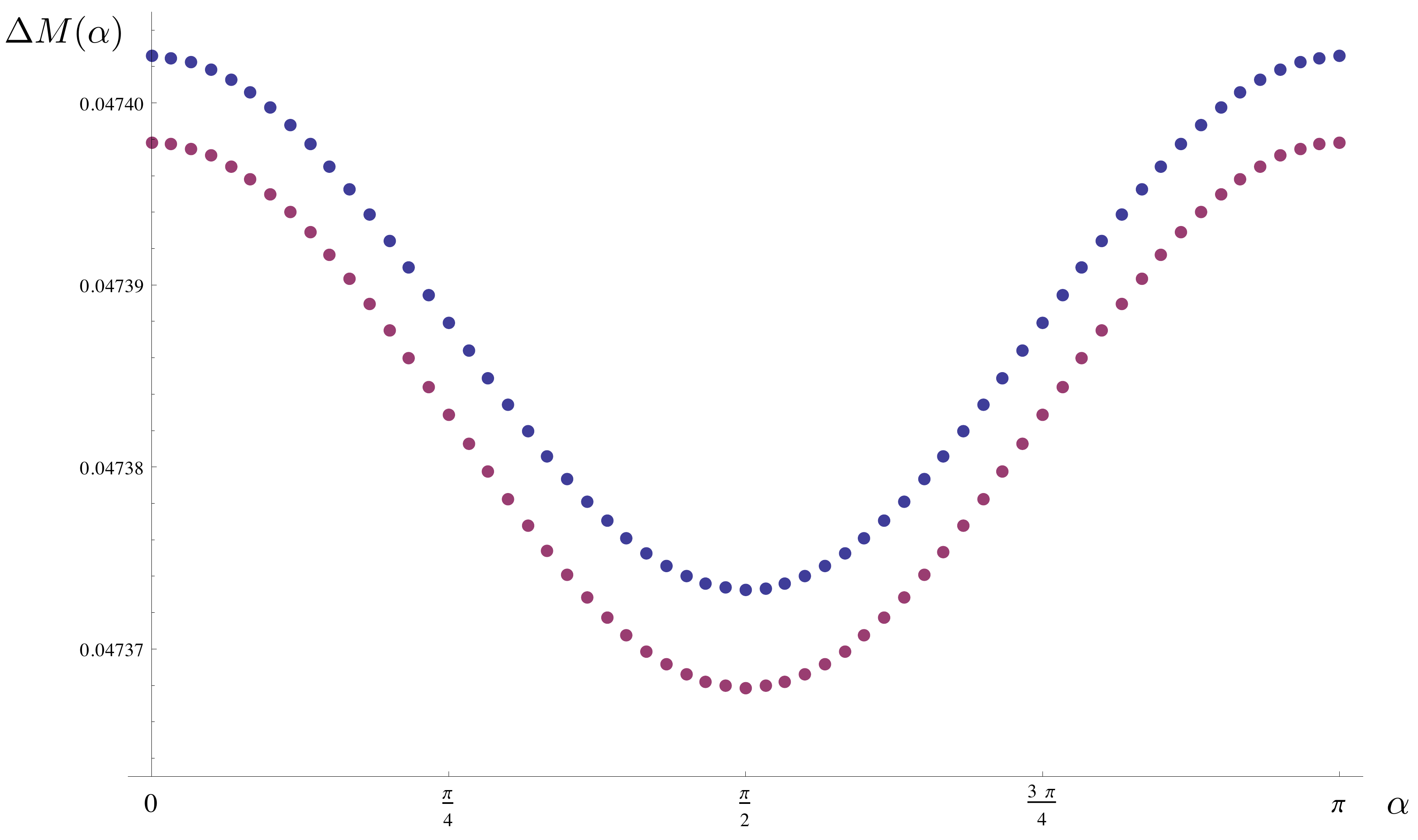}
 \caption{Plot of 60 values for $\Delta M$ in the interval $\alpha \in [0,\pi]$ for $\beta=0$ in two different scaling regimes. The blue curve is obtained for $\hat{k}=3.1667$, while the purple one is obtained for $\hat{k}=3.175$, which corresponds to $\epsilon = 0.36$ in~\eqref{scaling}. The values of the purple curves have been multiplied by $0.98$ times the ratio of the two $\Delta M$  for the collinear configurations in the two regimes.}
 \label{alpha_trend}
 \end{figure} 

On the contrary, if one keeps $\beta$ fixed and varies $\alpha$, the mass of the anti-supertube $\Delta M(\alpha, \beta)$ decreases as $\alpha$ starts increasing. This behavior is again independent on the particular choice for $\hat{k}$ in~\eqref{charges}, as shown in Figure~\ref{alpha_trend}. 

This proves that the collinear near-BPS microstate configuration obtained in Section~\ref{section_review} is classically unstable. Indeed, the initial microstate corresponds to $(\alpha,\beta)=0$ for the parametrization of Figure~\ref{parametrization} and Figure~\ref{alpha_trend} shows that $\Delta M$ decreases when $\beta=0$ and $\alpha$ starts increasing. The maximum relative difference in $\Delta M(\alpha,0)$ is 
\begin{equation}\label{energy_difference}
 \frac{\Delta M (0,0)-\Delta M (\pi/2,0)}{\Delta M(0,0)}\sim 6.1 \cdotp 10^{-4} 
 \end{equation}
 and in the scaling regime~\eqref{hamiltonian_scaling} this does not depend on the choice of $\hat{k}$. 
 
Of course it is interesting to ask whether this configuration will settle into another metastable minimum somewhere in the moduli space, or simply will keep decaying until it reaches a BPS minimum. In the slice that we explored there appears to be a minimum when $\alpha = \pi/2$ and $\beta=0$, but this does not imply that this minimum will be metastable. There could be other instability directions corresponding to other motions of the points in the moduli space. 

To prove that there will never be any metastable minimum one should investigate the full 12-dimensional moduli space\footnote{The moduli space of $N$ centers subject to the bubble equations and subtracting the center of mass motion is $2N-2$ dimensional.}, which seems computationally tricky to perform, especially because not all solutions to the bubble equations are free of closed time-like curves. Furthermore, if we eliminate the constraint that the total $J_L$ is zero, we can explicitly find a direction on the moduli space that leads to a scaling behavior, and as the configuration moves in that direction the energy of the anti-supertube approaches zero. Hence, we believe there is good reason to assume that metastable points in the moduli space are rare, if not altogether inexistent.


\section{The emission rates of non-extremal microstates}\label{section_strategy}
Figure~\ref{alpha_trend} shows that the collinear near-BPS solution is unstable in the six dimensional moduli space of solutions to the bubble equations~\eqref{bubbleequations} with the symmetry~\eqref{z2symmetry}. Classically, this instability would trigger a motion of the GH centers down the microstate throat and, in particular, Figure~\ref{alpha_trend} represents the potential energy that governs this motion. Quantum mechanically, this instability triggers a decay process towards extremality that causes the emission of radiation. From the thermodynamical point of view this is expected: a near-BPS black hole has a nonzero Hawking temperature and hence emits radiation according to (the five dimensional version of) Stephen's law. Our initial collinear near-BPS microstate would then decay into a series of different near-BPS microstates closer to extremality. We can interpret the symmetrical rotation in $\alpha$ of Figure~\ref{parametrization} as a possible initial decay channel of the collinear microstate and using $\Delta M(\alpha)$ shown in 
Figure~\ref{alpha_trend} we want to estimate the energy emitted per unit time $\Gamma$ into this particular decay channel at the beginning of the decay cascade.

We consider the initial decay process from the collinear microstate to the state given by the configuration $\alpha=\pi/2, \beta=0$ of Figure~\ref{parametrization}, which represents the minimum energy state in the slice considered in Figure~\ref{alpha_trend}. The emitted energy for this process is given by
\begin{equation}\label{ddm}
\ddm \equiv \Delta M (\alpha=0)-\Delta M (\alpha=\pi/2)
\end{equation}
and is only a small fraction of the initial $\Delta M (0)$ - see~\eqref{energy_difference}. It is hence correct to say that $\alpha=\pi/2$ is only an intermediate state in the decay cascade that brings the microstate towards extremality. 

To define an emission rate we need to know how much time the system needs to emit the energy $\ddm$. We suppose that this average time is of the same order of the characteristic time scale $\tau$ that, classically, governs the motion for small $\alpha$ as seen from an observer at infinity. Alternatively, $\tau$ can be seen as the necessary time to have a measurable displacement from the collinear configuration seen from infinity. 

In order to find $\tau$ we need to estimate the kinetic and the potential energies corresponding to the classical motion of the GH centers in the moduli space of solutions to the bubble equations~\eqref{bubbleequations}. We continue focusing on $\mathbf{Z}_2$ symmetric solutions~\eqref{z2symmetry}, and we need to find the small-$\alpha$ expansion of the potential energy shown in Figure~\ref{alpha_trend}, as well as the small velocity expansion of the kinetic energy of the bubbles, $\displaystyle d^2 E \over \displaystyle d \dot{\alpha}^2 $ . 

For small $\alpha$, the mass above extremality behaves as
\begin{equation}\label{potential_approximated}
\Delta M(\alpha)= \Delta M_0\, \left(1-c_2 \, \alpha^2\right)
\end{equation}
where $c_2$ is a coefficient that can be found numerically and whose dependence of the charges of the solution and of the scaling parameters of the solution is discussed in detail in Section~\ref{section_typicality}.

The strategy for computing the kinetic term is much more involved. The few obvious guesses about how to do it, involving for example treating one of the GH centers as a probe in the background sourced by the others, do not give sensible answers. The essential reason is that the energy of a bubble does not come only from the GH centers but also from the fluxes wrapping this bubble, and to compute the total energy brought about by the slow motion of the bubbles one has to compute the full energy of all the fluxes as well, and integrate the result over the full highly-warped spacetime. 

Our strategy is to rather use the fact that some bubbles are much smaller than others, and therefore the collective motion of a certain small bubble has the same energy as the motion of the black ring that undergoes a geometric transition to form the small bubble. One can then ``falsely compactify'' the solution to a four-dimensional one, by adding a small constant in the harmonic function describing the Gibbons-Hawking base, and compute the kinetic energy corresponding to the motion of the black ring. The final result is clearly independent on the small constant we are adding, and hence does not change when taking this constant to zero and recovering the asymptotically 4+1 dimensional solution.

Hence, the strategy we use can be summarized in the following recipe:
\begin{enumerate}
\item We compactify the three-charge solution~\eqref{initialsetup} to four dimensions along the fiber $\psi$ by adding a constant to the $V$ harmonic function.
\item We compute a Lagrangian for the motion in the $\alpha$-direction excising GH centers $g_1$ and $g_2$ from the background and replacing them with a singular black ring having the same mass $M_{br}$ and charges at the same distance $R\equiv r_{23}$ from the center, as explained in the physical interpretation of our solution in Appendix~\ref{appendix_physics}. The black ring in 4D is treated as a massive point particle that rotates in $\alpha$ in the background sourced by the other centers and under the effect of the potential in~\eqref{potential_approximated} generated by the anti-supertube probe closed to $g_6$.

\item We assume that the four-dimensional metric is not affected by the slow rotation in $\alpha$, and hence that all the possible corrections to the background fields caused by this motion are negligible. In addition, the electromagnetic interactions between the black ring and the background can be neglected when the rotation is slow. 
\end{enumerate}

The first hypothesis helps to avoid useless computations as it allows to consider just point particles instead of extended objects. The supergravity solution~\eqref{initialsetup} is asymptotically five-dimensional. 
The further compactification along $\psi$ requires some caution. Indeed one needs to modify the asymptotic behavior of the GH space from $\mathbb{R}^4$ to $\mathbb{R}^3\times S_1$ by introducing a constant $\delta V$ in the function $V$ in~\eqref{harmonic_functions}. The radius of the compactified $S_1$, $r\sim 1/\delta V^2$, can be thought of as a modulus of the solutions. In addition, one is also allowed to introduce constants $\delta K^I$ in the definitions of the harmonic functions $K^I$ in~\eqref{harmonic_functions}, which modify the bubble equations~\eqref{bubbleequations} - see~\cite{balasubramanian} for more details. While all these moduli are completely arbitrary, they only specify the asymptotics of our solution and do not affect the computation 
we are interested in. The phenomenon that we want to study takes place deep into the black-hole-like throat and hence is not affected by the details of the asymptotic fields of the background. Therefore we can compactify to four dimensions with no risk of ambiguity.

The second hypothesis is really the key point of our computation. We have verified that taking away centers $g_1$ and $g_2$ from our system does not substantially modify the solution to the bubble equations~\eqref{bubbleequations}. Using a hat to denote quantities computed without $g_1$ and $g_2$ we have $\widehat{r}_{56}\simeq r_{56}=r_{23}$, $\widehat{r}_{34} \simeq r_{34}$, up to corrections of order $0.1\%$. Most importantly, we verified numerically that the probe anti-supertube still has a metastable point close to $g_6$ and the behavior under rotation is similar to that of the complete system:
\begin{align}
\Delta\widehat{M}(\alpha)=\Delta\widehat{M}_0(1-\widehat{c}_2\alpha^2)\quad \quad \Delta\widehat{M}_0 \simeq \Delta M_0, \quad \quad \widehat{c}_2 \simeq \frac{c_2}{2}\label{potentialexcised}
\end{align}
where $c_2$ was introduced in~\eqref{potential_approximated}\footnote{The factor in the relation between $\widehat{c}_2$ and $c_2$ comes from the fact that the energy reduction caused by the motion of the black ring alone is half of that caused by the movement of both the black ring and the bubbling black ring given by the points $g_6$ and $g_7$. Similarly, the kinetic energy corresponding to the motion of the black ring is half of that corresponding to the motion of both the black ring and the GH points $g_6$ and $g_7$.}. This means that we can excise the GH centers $g_1$ and $g_2$ from the whole solution and replace them by a BPS black ring whose charges are determined by the sum of the residues in the $K,L$ and $M$ harmonic functions of the points we excised. This is the inverse of the bubbling black ring transition described in \cite{Bena_bubbling_supertubes}, and it is not hard to check that the distance of the black ring from the central blob given by $g_3,g_4$ and $g_5$ is essentially the same as the original distance between the blob and the excised GH centers. Since the points $g_6$ and $g_7$ have not been excised, we can put the anti-supertube probe with charges~\eqref{chargesupertube} at the metastable location close to~$g_6$. 

Our strategy is to compute the kinetic term corresponding to the rotation of the black ring center in $\alpha$ (described in Figure~\ref{parametrization}), by treating the black ring center as a probe in the background sourced by the other GH centers. 

Finally, the third hypothesis is reliable because $\alpha$ and its derivatives with respect to time are small and hence modifications to the four dimensional metric become higher-order corrections. Therefore we can safely use the four-dimensional metric generated by the collinear solution to estimate the kinetic term. Note that since the background is kept fixed to first order in $\alpha$, the electromagnetic interactions between the point-like black ring and the background are negligible.


\subsection{The decay time and the emission rate}
We compactify the eleven-dimensional supergravity solution~\eqref{background} on $T_6$ and $\psi$ to a four-dimensional solution whose metric is
\begin{align}
ds^2_4&=J_4^{-\frac{1}{2}}(dt +\omega)^2+J_4^{\frac{1}{2}}[dr^2+r^2(d\alpha^2+\sin^2\alpha\, d\phi^2)] \nonumber \\
J_4&=Z_1 Z_2 Z_3 V-\mu^2V^2 \label{4dbackground}
\end{align}
and the four-dimensional dilaton is constant \cite{Berglund}. Note that the positivity of $J_4$ is one of the fundamental requirements for the construction of the solution~\cite{Bena:microstates}, and comes from the absence of closed time-like curves in the eleven-dimensional geometry.

The GH centers $g_1$ and $g_2$ have been excised from the background~\eqref{4dbackground} and we denote with a hat all the quantities computed using only the centers $g_3,...,g_7$ with parameters as in~\eqref{charges}. The black ring corresponds to a point-like particle at a distance 
\begin{equation}\label{R_definition}
R=r_{23}=\hat{r}_{45}
\end{equation}
from $g_4$ substituting $g_1$ and $g_2$ in Figure~\ref{figurecolinear}. Its mass $M_{br}$ corresponds to the mass associated with $g_1$ and $g_2$ and is nothing but the mass of the black ring microstate that these centers represent, as explained in Appendix~\ref{appendix_physics}. The latter was found in~\cite{Bena:mergers} 
\begin{align}
M_{br}&=Q^1_{br}+Q^2_{br}+Q^3_{br} \quad \quad \quad Q^I_{br}=C^{IJK}d^Jf^K \label{mass_supertube}
\end{align}
where we have introduced the parameters
\begin{align}
d^I&=2\left(k_1^I+k_2^I\right) \quad  \quad f^I=6k_0^I+\left(1+\frac{1}{v_1}\right)k_1^I+\left(1-\frac{1}{v_2}\right)k_2^I \quad \quad  k_0^I=\frac{1}{3}\left(k_3^I+k_4^I+k_5^I \right)
\end{align}
If we then let the system rotate along $\alpha$, the re-inserted point particle interacts with the potential~\eqref{potentialexcised}. To compute the full Lagrangian we need to find the kinetic term corresponding to the slow motion of the black ring in the moduli space, and for this we can use the classical GR action:
\begin{equation}
S=-M_{br}\int \sqrt{-\hat{g}_{\mu \nu}\frac{dx^{\nu}}{dt}\frac{dx^{\mu}}{dt}}=-M_{br}\int \sqrt{\tilde{J}_4^{-\frac{1}{2}}-\tilde{J}_4^{\frac{1}{2}} \,R^2\dot{\alpha}^2}\sim \int\frac{1}{2} M_{br}\tilde{J}_4^{\frac{3}{4}}\, R^2 \dot{\alpha}^2+const
\end{equation}
where we have used the time $t$ measured by an observer at infinity to parameterize the worldline. Note that a tilde above $J_4$ means that this quantity is evaluated at the location of the black ring, namely at a distance $R$ from the center $g_4$.

Thus, the full Lagrangian corresponding to the motion of the GH centers that triggers the decay of the metastable supertube 
\begin{equation}
\mathcal{L}= \frac{1}{2}M_{br}\tilde{J}_4^{\frac{3}{4}} R^2 \dot{\alpha}^2-\Delta\widehat{M}(\alpha)
\end{equation}
 and the associated equation of motion to first order in $\alpha$ is
\begin{equation}
M_{br}\tilde{J}_4^{\frac{3}{4}} R^2 \ddot{\alpha}-2\widehat{c}_2 \Delta M_0 \alpha =0
\end{equation}
Equation~\eqref{potentialexcised} allows us to approximate $2\widehat{c_2}\sim c_2$ and hence the characteristic time scale of this differential equation is
\begin{equation}\label{tau}
\tau=\sqrt{\frac{M_{br}\tilde{J}_4^{\frac{3}{4}} R^2}{\Delta M_0 c_2}}
\end{equation}

This equation is the main result of our paper. Since $\Delta M_0$ parameterizes the initial energy of this solution above the BPS bound we see that the closer our solution is to the BPS bound the bigger $\tau$ is. 

 Using~\eqref{ddm} and~\eqref{tau} following the arguments presented in Section~\ref{section_strategy} we define the emission rate of our microstate in the $\alpha$-channel to be
 \begin{equation}\label{gamma}
 \Gamma=\frac{\ddm}{\tau}
 \end{equation}
 In the next section we study how $\Gamma$ scales under a scaling of the conserved charges $Q_I$ and of the distances in the ${\mathbb R}^3$ base of the Gibbons-Hawking space underlying the BPS solution. These results will be used to check whether this particular kind of emission from the initial collinear microstate is typical in the thermodynamical ensemble.


\section{Typicality of the $\alpha$ decay channel}\label{section_typicality}

\subsection{Scaling properties of the $\alpha$-emission rate}\label{section_scaling}

In this section we study how $\Gamma$ behaves under two different types of scaling of the background. The results of this section will be used in the next one to compare $\Gamma$ with the thermal emission rate found by the authors of~\cite{callan} for a D1-D5-P non-extremal black hole and thus gain information about the typicality of the $\alpha$-decay channel studied in the previous sections. 

We are interested in two separate scalings that involve the parameters of our microstate.
The first one corresponds to the scaling of the depth of the throat of our microstate~\eqref{scaling} while keeping the charges and fluxes essentially fixed:
\begin{align}\label{epsilon_scaling}
r_{ij}\rightarrow \epsilon \, r_{ij} 
\end{align}
The second one corresponds to scaling all the magnetic fluxes in~\eqref{charges} 
\begin{align}\label{xi_scaling}
\quad \quad \quad k\rightarrow \xi k
\end{align}
and modifies the charges and the mass as~\eqref{bhchargesfixed}:
\begin{align}
Q_I \rightarrow \xi^2 Q_I \quad \quad \quad M\rightarrow \xi^2 M \label{chargescaling}
\end{align}
  
Note that $\delta M_0$ is considered as a free parameter of the system and it does not scale. This quantity is the energy \eqref{metastable0value}  brought about by placing an anti-supertube probe at metastable point close to $g_6$  and can be kept fixed while performing the scalings above by suitably tuning the anti-supertube charges~\eqref{chargesupertube}.
For the particular anti-supertube we are using, we only need to tune the probe charge $q_2$ and keep $q_1$ and $d_3$ fixed as in~\eqref{chargesupertube}. This is because the charge $q_2$ has opposite sign with respect to the corresponding background charge and it is responsible for supersymmetry breaking and metastability. 

To study how $\tau$ in~\eqref{tau} scales with $\epsilon$ in~\eqref{epsilon_scaling} one can use equations~\eqref{hamiltonian_scaling} and~\eqref{potential_approximated} to deduce that $c_2$ in~\eqref{tau} does not transform, which is also shown in Figure~\ref{alpha_trend}. Then using the formulas in Appendix~\ref{appendix_background} it is easy to determine that $\tilde{J}_4 \rightarrow \epsilon^{-3}\tilde{J}_4$, $R^2\rightarrow \epsilon^2 R^2$ and hence
\begin{equation}\label{tau_scaling_epsilon}
\tau \rightarrow \epsilon^{-\frac{1}{2}} \tau
\end{equation}
As pointed out in~\cite{Bena:supertube_microstate}, the physical importance of $\epsilon$ is to scale the (metric) length $L_{MS}$ of the microstate throat as\footnote{The metric length of the microstate~\cite{Bena:supertube_microstate} is given by  
$ \displaystyle 
L_{MS}=\int_{z_7}^{z_{neck}}V^{\frac{1}{2}}(Z_1Z_2Z_3)^{\frac{1}{6}}dz,
$
where $z_{neck}$ is a suitable cutoff.} 
\begin{equation}\label{throatscaling}
L_{MS}\rightarrow \epsilon^{-1}L_{MS}
\end{equation}
Because of equations~\eqref{tau_scaling_epsilon} and~\eqref{throatscaling} we see that the decay time corresponding to the rotation $\alpha$ becomes longer as the length of the throat becomes longer; thus, the closer to BPS the configuration is the slower it decays.

To study how $\tau$ in~\eqref{tau} scales with $\xi$ in~\eqref{xi_scaling} it is important to observe that after the scaling the bubble equations~\eqref{bubbleequations} are exactly solved by $ r_{ij} \rightarrow \xi^2 r_{ij} $. Then it is easy to verify that
\begin{align}
M_{br} \rightarrow \xi^2 M_{br} \quad \quad \quad R^2 \rightarrow \xi^4 R^2 \quad \quad \quad \tilde{J}_4 \rightarrow \xi^{-2} \tilde{J}_4
\end{align}
Unlike for the $\epsilon$-scaling, we could not evaluate analytically the scaling properties of $c_2$ in~\eqref{tau}. To infer them numerically one can start from a ($\xi=1$) solution with charges given in ~\eqref{chargesmacroscopic}, and in order to keep $\Delta M_0$ in~\eqref{metastable0value} fixed while varying $\xi$ one needs to change the $q_2$ charge of the anti-supertube probe in~\eqref{chargesupertube}:
\begin{center}
\begin{tabular}{|c|c|}
\hline
$\xi$ & $q_2$ \\
\hline
$1$ & $-50$ \\
$1.2$ &$ -63.522$ \\
$2$ & $-117.92 $\\
$3$ & $-186.355$ \\
$4$ & $-254.958 $\\
\hline
\end{tabular}
\end{center}
By repeating for each value of $\xi$ the evaluation that leads to the potential shown in Fig. \ref{alpha_trend}, one finds that 
\begin{equation}\label{c2_xi_scaling}
c_2\rightarrow \xi^{-\frac{1}{2}}c_2
\end{equation}
and therefore the overall scaling of $\tau$ with $\xi$ under~\eqref{xi_scaling} is:
\begin{equation}\label{tau_scaling_xi}
\tau \rightarrow \xi^{\frac{5}{2}} \tau
\end{equation}

Finding the scaling properties of $\ddm$ is much easier. This quantity does not scale with $\epsilon$. Indeed, the probe hamiltonian~\eqref{hamiltonian} does scale with $\epsilon$, as described in equation~\eqref{hamiltonian_scaling}, but as we tune the probe charge to keep $\Delta M_0$ constant it turns out that $\ddm$ also remains constant. This does not happen for the $\xi$-scaling~\eqref{xi_scaling} and as there is no analytical formula for $\ddm$ it is necessary to perform another numerical interpolation. Tuning the probe charge $q_2$ as before we find
\begin{equation}\label{ddm_scaling_xi}
\ddm \rightarrow \xi^{-0.8}\ddm
\end{equation}

Finally, using the definition~\eqref{gamma} with~\eqref{tau_scaling_epsilon}, \eqref{tau_scaling_xi} and~\eqref{ddm_scaling_xi} one determines the scaling properties of $\Gamma$ under~\eqref{epsilon_scaling} and ~\eqref{xi_scaling}:
\begin{align}
\Gamma \rightarrow \epsilon^{\frac{1}{2}} \Gamma \quad \quad \quad \Gamma \rightarrow \xi^{-3.3}\Gamma \label{gamma_scaling}
\end{align}
In the next section this result will be used to compare the emission rate $\Gamma$ in the $\alpha$-channel with the emission rate of the thermodynamical ensemble to check whether this decay is typical.


\subsection{How typical is the $\alpha$ decay channel ?}\label{subsection_typicality}
In this section we want to compare the emission process of our microstate with the thermal emission of a near-BPS five-dimensional Reissner-Nordstr\"om black hole. In particular, we want to check whether the $\alpha$ channel emission has the features that one expects from thermodynamics.

 We have three fundamental pieces of data about the radiation emission of the thermodynamical ensemble coming from general relativity and brane technology. The first is the computation~\cite{callan} of the emission rate for a near-BPS three-charge five dimensional black hole in the D1-D5-P frame, where a tiny amount $N_L$ of left-moving momentum is inserted on a string of length $N_1 N_5$ with $N_R>>N_L$ right-moving momentum, so to break supersymmetry. The energy emission rate for closed strings was found to be
\begin{equation}\label{thermal_emission}
\Gamma_{th}\sim \sqrt{Q_1Q_2Q_3}T_H^5
\end{equation}
where $T_H$ is the Hawking temperature of a five-dimensional Reissner-Nordstr\"om black hole computed from its surface gravity:
\begin{equation}\label{hawking}
T_H \sim \frac{1}{R_e}\sqrt{\frac{\Delta M_0}{M}}
\end{equation}
Here $R_e$ and $M$ are the horizon radius and mass of the black hole, and $\Delta M_0$ is 
the mass above extremality, which is assumed to be much smaller than $M$.
 Given that the horizon area for this class of black holes is proportional to $A\propto \sqrt{Q_1 Q_2 Q_3}$ we see that~\eqref{thermal_emission} is simply Stephen's law in five dimensions. 

The second piece of data for the comparison of energy emission rates is Wien's law, which follows from the five-dimensional version of Planck's law and has the same form in four and five dimensions:
\begin{equation}\label{wien_law}
\nu_{max, th} \sim T_H
\end{equation}
where $\nu_{max, th}$ is the peak-frequency of energy emission from the ensemble and $T_H$ is the black hole temperature~\eqref{hawking}.

{The third and final piece of data is the difference $\Delta L$ between the depth of the microstate throat $L_{ms}$ and the throat $L_{bh}$ of the black hole corresponding to the microstate\footnote{This is a near-BPS five-dimensional Cveti\v c-Youm black hole~\cite{Cvetic:1996xz}.}. Since supersymmetric black holes have infinite $L_{bh}$, this comparison is meaningful only for non-supersymmetric black holes. In~\cite{Bena:supertube_microstate} it was shown that the for class of near-BPS microstates we discuss one can arrange $L_{ms}$ to be arbitrarily  larger or smaller than $L_{bh}$. However, we expect that typical microstate geometries of the black hole will have $L_{ms}$ comparable to $L_{bh}$, and hence $\Delta L \approx 0$. In~\cite{Bena:supertube_microstate} $\Delta L$ was found to be
\begin{equation}\label{DeltaL}
\Delta L = L_{bh}-L_{ms} = \rho_{neck} \ln \left( 2\frac{\rho_{ms}}{\rho_{bh}} \right)
\end{equation}
The parameter $\rho_{ms}$ in~\eqref{DeltaL} is given by
\begin{equation}
\rho_{ms}= 2 \sqrt{R}
\end{equation}
where $R$ is the distance of the outermost GH center~\eqref{R_definition}, while $\rho_{bh}$ is given by:
\begin{equation}\label{rhobh}
\rho_{bh}^2= \sqrt{\frac{8 \Delta M}{\frac{1}{Q_1}+\frac{1}{Q_2}+\frac{1}{Q_3}}}
\end{equation}
which is the horizon radius of the corresponding non-extremal black hole. The parameter $\rho_{neck}$ in~\eqref{DeltaL} corresponds to a certain cutoff needed to measure the throat lengths, but is irrelevant if when one imposes $\Delta L=0$, which implies
\begin{equation}\label{typicalradius}
\rho_{bh} = 2\rho_{ms}
\end{equation}

The scaling properties of $\rho_{ms}$ under~\eqref{epsilon_scaling} and~\eqref{xi_scaling} are easily found from the results of Section~\ref{section_scaling}:
\begin{equation}\label{scaling_rhoms}
\rho_{ms} \rightarrow \epsilon^{\frac{1}{2}}\rho_{ms} \quad \quad \quad \rho_{ms} \rightarrow \xi \rho_{ms}
\end{equation}}
It is straightforward to determine how~\eqref{thermal_emission}, \eqref{hawking} and~\eqref{rhobh} scale with $\xi$ under~\eqref{chargescaling}:
\begin{align}
T_H \rightarrow \xi^{-2}T_H, \quad \quad \Gamma_{th}\rightarrow \xi^{-7} \Gamma_{th}, \quad \quad \rho_{bh} \rightarrow \xi^{\frac{1}{2}} \rho_{bh}  \label{thermal_scaling}
\end{align}
The thermal quantities do not scale with $\epsilon$. 

Given the three equations describing the thermal emission of the ensemble~\eqref{thermal_emission} and the size of the microstate,~\eqref{wien_law} and~\eqref{typicalradius}, we can argue that a given decay process of a microstate into a particular channel is typical if its $\Gamma$, $\nu_{\max}$ and $\rho_{ms}$ match those given by these equations.
The only missing information about our decay channel is $\nu_{max}$. We can argue that the energy emitted during the decay of the nonextremal microstate geometry is given by the difference between the highest and the lowest values of the mass above extremality during the rotation in $\alpha$:
\begin{equation}
\nu_{max} = \ddm
\end{equation}

Given the scaling properties of our solutions~\eqref{epsilon_scaling},~\eqref{xi_scaling},\eqref{tau_scaling_epsilon}, \eqref{tau_scaling_xi} and~\eqref{scaling_rhoms} we can start from the initial solution (with charges given in equation~\eqref{chargesmacroscopic})
and check whether there is any value of $\epsilon$ and $\xi$ (or alternatively of $Q^\prime_1,Q^\prime_2, Q^\prime_3$ and $\rho_{ms}^\prime$) for which the decay of our solutions matches the thermal decay:
\begin{align}
\Gamma_{th}^\prime = \Gamma^\prime \quad \quad 
\nu_{max,th}^\prime = \nu_{max}^\prime \quad \quad 
\rho_{bh}^\prime = 2\rho_{ms}^\prime \label{thermal_condition}
\end{align}

Unfortunately, this is not possible, which implies that the solution we started from and the families of non-extremal microstate solutions obtained by scaling its depth and fluxes via \eqref{epsilon_scaling} and \eqref{xi_scaling} are not typical. This is not surprising - after all, we started from a very specific seven-center solution that has a lot of symmetry, and a very large ratio between certain inter-center distances, and we examined a non-extremal microstate geometry obtained by adding a certain type of anti-supertube to this solution. It would have been in fact much more surprising if this decay process had been thermal-like.

{It is also possible to parameterizes  the departure of the non-extremal microstate we consider from typicality by introducing a quantity, $\beta$, that can be thought of as modifying the  estimation of $\tau$ of Section~\ref{section_strategy}\footnote{The choice of multiplying  $\tau$ is not arbitrary: we fully trust the assumptions that enter in the computation of $\tau$ of Section~\ref{section_strategy}, but given that we have not investigated the motion of the GH centers in the full six-dimensional moduli space, we do not know whether other decay directions exist where the decay time is faster or shorter. For example, if one would rather rotate the segment formed by the central points of the solutions, $g_3, g_4$ and $g_5$, the inertia momentum would be much smaller than the one corresponding to rotating the black ring centers, and hence the decay time would be much faster.}. If one multiplies $\tau$ by $\beta$ in all the equations above, the system~\eqref{thermal_condition} can always be solved for some $\xi_{th}, \epsilon_{th}, \beta_{th}$. Plugging in the numbers we find that 
\begin{equation}\label{beta_result}
\beta_{th} \sim 1.8 \cdotp 10^{5}
\end{equation}

This therefore gives an estimate of the departure from typicality of the microstate we have considered. The final result $\beta_{th}>>1$ has two possible implications that are not mutually exclusive. On one hand, it  can imply that a typical microstate should have a $\Gamma$ that is much smaller than the one of our emission process, or a much larger characteristic decay time $\tau$.

{
This could be for example realized by considering non-extremal microstates that have more centers, and whose moment of inertia for the rotation in the direction that lowers the energy of the microstate is much bigger than in our solution. On the other hand~\eqref{beta_result} suggests also that the emitted energy $\ddm$ should be much smaller than the one we found. This is not helped at all by considering microstate geometries that have more centers, because the potentials of these geometries will be generically less abrupt. Hence, to decrease the emitted energy one should rather consider creating non-extremal microstates by adding supertubes to BPS microstates with simple topology (like the superstrata of \cite{superstrata-counting}). It would be interesting to investigate which of the two options is the best for producing more typical non-extremal microstate geometries. }


\section{Conclusions and outlook}\label{section_conclusions}
We analyzed the near-BPS smooth collinear microstate built in~\cite{Bena:supertube_microstate} by adding a probe anti-supertube at a metastable minimum of its potential in the background of the smooth microstate for a BMPV black hole with large horizon area~\cite{Bena:mergers}. We proved that this microstate is unstable by showing that the potential energy of the supertube $\Delta M$ decreases if the GH centers of the microstate solutions rotate with respect to the axis of the initial collinear configuration.

We also estimated in Section~\ref{subsection_typicality} the typical instability time and the emission rate corresponding to this decay, and compared the results to those expected from thermodynamics. We used three known thermodynamic quantities: the emitted energy rate, the peak frequency and the typical radius of microstates, and we showed that they cannot be all matched by the features of the decay channel we have found. We also estimated the mismatch to be of order $10^5$, which can be thought of as a measure of the departure from typicality of the nonextremal black hole microstate solution we started from. 

It is clearly important to construct non-extremal microstate geometries that have a more typical decay. One way to do this is to  build near-BPS microstates with more that seven centers, or whose centers are more evenly spaced as the ones in our solutions (which are at distances whose aspect ratio is of order 1000). Indeed, the seven-centers solution analyzed in this paper represents somehow the minimal interesting model that one can build~\cite{Bena:mergers}. Furthermore, the moment of inertia corresponding to the motion that destabilizes our solution is very large as it corresponds to moving an entire bubbled black ring that has about one third of the total mass of the microstate. Finding a more sophisticated solution, though more involved, will allow us to look for analogous decay patterns that have a parametrically smaller moment of inertia or parametrically larger $\ddm$, such that  and decay time that is fast-enough to be in the typical range. 

From a more general perspective, our result confirms the intuition of~\cite{Chowdhury-Mathur}, that most of the microstate solutions of non-extremal black holes should be unstable, and hence the dynamics of these black holes will display a chaotic behavior, corresponding to microstates being formed and immediately decaying into other microstates, which in their turn decay very fast. It would be interesting to see if our solutions can be used to shed light on some of the features of the chaotic behavior of non-extremal black holes discussed in~\cite{Shenker}.

Besides its implications for black hole physics, our result may also have important consequences for the program of uplifting AdS vacua obtained from generic flux compactifications to obtain de Sitter space in string theory. There is a direct analogy between the uplift of AdS solutions to de Sitter by adding antibranes~\cite{KKLT, Kachru} and the uplifting of BPS microstate geometries to microstates of non-BPS black holes by adding anti-supertubes~\cite{Bena:supertube_microstate}. In both constructions one used the action of the probe antibranes to argue that they have metastable vacua. However, our investigation reveals that the result of the probe calculation can be misleading, and that the metastable supertube can be in fact destabilized by the motion in the moduli space of the underlying geometry. However, unlike microstate geometries, flux compactifications usually come with all the moduli stabilized. Nevertheless, it is possible that even stabilized moduli can be destabilized, especially when their mass is very low. It would be interesting to understand whether this happens when investigating antibranes~\cite{Kachru, Bena:anti_D3, Bena:giant_tachyon} in the Klebanov-Strassler warped deformed conifold solution~\cite{Klebanov}.

\section*{Acknowledgments} 
We are grateful to David Turton and Nick Warner for useful discussions and suggestions. This work is supported in part  by  the ERC Starting Grant 240210 \textit{String}-QCD-BH, by the John Templeton Foundation Grant 48222 and by a grant from the Foundational Questions Institute (FQXi) Fund, a donor advised fund of the Silicon Valley Community Foundation on the basis of proposal FQXi-RFP3-1321 (this grant was administered by Theiss Research).

\appendix

\section{Construction of the three-charge smooth background}\label{appendix_background}
 
 In this appendix we review the construction of smooth three-charge black hole microstate solutions. Some aspects of physics of these solutions are discussed in the next appendix, and more details can be found in \cite{Bena:microstates,Gibbons:2013tqa}.

 The smooth three charge solution preserves $\mathcal{N}=4$ supercharges in 11D supergravity compactified on  three tori. The metric and the three-form potential are given by:
 \begin{align}
 ds^2_{11} &=-(Z_1Z_2Z_3)^{-\frac{2}{3}}(dt+k)^2+(Z_1Z_2Z_3)^{\frac{1}{3}}ds_4^2+(Z_1Z_2Z_3)^{\frac{1}{3}} 
\sum_{i=1}^{3}\frac{dx_{4+i}^2+dx_{5+i}^2}{Z_i} \nonumber \\
 A^{(3)} &=A^1 \wedge dx^5 \wedge dx^6 +A^2\wedge dx^7 \wedge dx^8 +A^3\wedge dx^9 \wedge dx^{10} \label{background}
 \end{align}
 where $k$ is the angular momentum vector, $Z_i$ are the three warp factors associated to the electric conserved charges and $dx_{4+i}^2+dx_{5+i}^2$ for $i=1,2,3$ is the standard metric on a torus. The metric of the base space of this solution, $ds^2_4$, is chosen to be a multi-center Gibbons-Hawking/Taub-NUT metric:
  \begin{equation}\label{GHmetric}
  ds_4^2=V^{-1}(d\psi +\vec{A}\, \cdotp d\vec{y})^2+V(dy_1^2+dy_2^2+dy_3^2)
  \end{equation}
 where
 \begin{equation}\label{A}
\vec{\nabla} \times \vec{A}=\vec{\nabla} V
\end{equation}
 and the Taub-NUT fiber $\psi$  has period $4\pi$. All the functions  that appear in the background~(\ref{background}) and in~\eqref{GHmetric} depend on $(y_1, y_2, y_3)$, and the full solution is completely determined by specifying four harmonic functions:
 \begin{align}
 V=\sum_{j=1}^{N} \frac{v_j}{r_j} \quad \quad K^{I}=\sum_{j=1}^{N} \frac{k_j^I}{r_j}\quad  \quad I=1,...3 \quad \quad r_j=|\vec{y}-\vec{g}_j| \label{harmonicfunctions}
 \end{align}
  where $N$ is the number of GH centers located at $\vec{g}_i$ and $(v_j,k^I_j)$ are parameters to specify. Notice that one can potentially add some constants $\delta V$ and $\delta K^I$ to the functions in~\eqref{harmonicfunctions} so that the GH space asymptotes $\mathbb{R}^3\times S_1$ and the $\delta K^I$ generate some Wilson lines for the three form in~\eqref{background}. As we want the GH space to asymptote $\mathbb{R}^4$ these constants are taken to be zero.
  
 When $v_j \in \mathbb{Z}$, the GH centers become benign orbifold singularities. The geometry in~\eqref{GHmetric} asymptotes to flat $\mathbb{R}^4$ if one also requires
  \begin{equation}\label{constraint1}
  \sum_{j=1}^N v_j=1
  \end{equation}
The GH metric~\eqref{GHmetric} is then ambipolar, meaning that its signature switches from $+4$ to $-4$. However, as shown in \cite{Bena_bubbling_supertubes,Berglund}, this is not a problem for the full solution~\eqref{background}, which is everywhere smooth and Lorentzian.

The warp factors $Z_I$ of the solution are
 \begin{equation}\label{warpfactors}
 Z_I=L_I+\frac{1}{2}C_{IJK}\frac{K^IK^J}{V}
 \end{equation}
where the $L_I$ are harmonic functions in the GH space. Requiring the $Z_I$ to be smooth at the GH centers and fixing their asymptotic value to $1$ imply that
   \begin{equation}
   \label{smooth1}
    L_I = 1- \frac{1}{2}C_{IJK}\sum_{j=1}^N \frac{k_j^Ik_j^K}{v_j}\frac{1}{r_j} 
    \end{equation}
   where $C_{IJK}\equiv |\varepsilon_{IJK}|$. The BPS solution for the angular momentum vector $k$ is written as  
  \begin{equation}\label{momentum_k}
 \vec{k}=\mu(d\psi +\vec{A})+\vec{\omega}
 \end{equation}  
 where $A$ is defined in~\eqref{A} and $\mu$ is given by
  \begin{equation}\label{mu}
 \mu=\frac{C_{IJK}K^IK^JK^K}{6V^2}+\frac{K^IL_I}{2V}+M
 \end{equation}  
 with $M$ another harmonic function. The vanishing of $\mu$ at the GH centers determines $M$ as
\begin{equation}
\label{smooth2}
    M = m_0 +\frac{1}{12} C_{IJK} \sum_{j=1}^N \frac{k_j^Ik_j^Jk_j^K}{v_j^2}\frac{1}{r_j}
    \end{equation}
  where $m_0$ is a constant whose value is found requiring that $k$ vanishes at infinity:
    \begin{equation}
    m_0=-\frac{1}{2}\sum_{j=1}^N\sum_{I}k_j^I
    \end{equation}
 The last form to define in~\eqref{momentum_k} is $\vec{\omega}$, which is given by:
 \begin{align}
 \vec{\omega}=\frac{1}{24}C_{IJK}\sum_{i,j=1}^N v_i v_j \Pi_{ij}^I \Pi_{ij}^J \Pi_{ij}^K \vec{\omega}_{ij} \quad \quad \Pi_{ij}^{I}\equiv \frac{k^I_j}{v_j}-\frac{k^I_i}{v_i} \label{omega}
 \end{align}
 where, choosing a coordinate system with $\vec{y}^i=(0,0,a)$, $\vec{y}^j=(0,0,b)$ with $a>b$ and defining $\tan \phi=y_2/y_1$, one has:
 \begin{equation}
 \vec{\omega}_{ij}=-\frac{y_2^2+y_1^2+(y_3-a+r_i)(y_3-b-r_j)}{(a-b)r_ir_j}d\phi
 \end{equation}
To avoid the existence of closed-timelike-curves (CTCs) it is necessary that 
\begin{equation}
Z_1Z_2Z_3V-\mu^2 V^2 \geq 0
\end{equation}
holds everywhere in the GH space. 

Furthermore, to avoid Dirac-Misner strings, the solution must satisfy the \textit{bubble equations} that constrain the distances between the GH centers \cite{bates_denef, Bena_bubbling_supertubes, Berglund}: 
  \begin{equation}\label{bubbleequations2}
  \sum_{j=1\, j\neq i}^{N} \Pi_{ij}^{(1)}\Pi_{ij}^{(2)}\Pi_{ij}^{(3)}\frac{v_iv_j}{r_{ij}}=-2\left(m_0\,v_i+\frac{1}{2}\sum_{I=1}^{3}k_i^I\right)
  \end{equation}
whith $\Pi_{ij}^I$ as in~\eqref{omega}. 
Only $N-1$ out of $N$ bubble equations are independent: indeed summing the LHS of~\eqref{bubbleequations2} over $i$ one gets zero as the $\Pi_{ij}^I$  are anti-symmetric in $ij$. For smooth microstate solutions constructed using GH centers only~(\ref{smooth1},\ref{smooth2}) this condition is equivalent to the vanishing of $\mu$ in~\eqref{mu} at every GH center~\cite{Bena:mergers}.

Expanding the warp factors $Z_I$ in~\eqref{warpfactors} and the momentum vector of~\eqref{momentum_k} it is possible to read the three electric charges $Q_I$ and the two angular momenta $J_1$ and $J_2$ preserved by this background. In particular once the parameters $(v_j, k_j^I)$ have been specified the charges $Q_I$, and the sum of the angular momenta are:
\begin{align}
Q^I&=-2C^{IJK}\sum_{j=1}^Nv_j^{-1}\tilde{k}^J_j \tilde{k}^K_j \nonumber\\
J_1+J_2 &=\frac{4}{3}C^{IJK}\sum_{j=1}^N v_j^{-2}\tilde{k}^I_j\tilde{k}^J_j \tilde{k}^K_j   \label{bhchargesfixed}
\end{align}
where
\begin{align}
\tilde{k}^I_j \equiv k^I_j-v_j\sum_s k^I_s
\end{align}
The expression for the difference of the angular momenta depends also on the positions, $\vec{g}_i$, of the GH centers~\cite{Berglund}}:
\begin{align}
J_1-J_2 = 8|\vec{D}| \quad \quad \vec{D}=\sum_{I=1}^{3}\sum_{j=1}^{N}\tilde{k}^{I}_j\vec{g}_j \label{Jl}
\end{align}
Finally the electric three-form in~\eqref{background} is specified by
\begin{equation}
dA^{I}=\Theta^I -d\left(\frac{dt+k}{Z_I}\right)
\end{equation}
where $\Theta^I$ are the dipole field strengths
\begin{equation}\label{dipole_forms}
\Theta^I=-\sum_{a=1}^3 [\partial_a(V^{-1}K^I)][(d\psi+A)\wedge dy^a+\frac{1}{2}V\varepsilon_{abc} \, dy^b\wedge dy^c]
\end{equation}
In the formula above $A$ is the three form computed in~\eqref{A}. In the next section we give a physical interpretation of some bubble solution and explain the role of the dipole field strengths.

\section{Some physical properties of bubbled solutions}\label{appendix_physics}
In this section we explain how to use the technique presented in Appendix~\ref{appendix_background} to build microstates of BPS black holes and black rings. The solution~\eqref{background} is a \textit{bubbling} one, which indicates the fact that the black hole or black ring solutions, that one would get if the metric $ds^2_4$ in~\eqref{background} were that of flat $\mathbb{R}^4$, are replaced by horizonless solutions that we built using a Gibbons-Hawking metric~\eqref{GHmetric} that has the same asymptotics but nontrivial topology\footnote{This topology is easy to see: the periodic coordinate $\psi$ of~\eqref{GHmetric} is fibered over a curve connecting two GH centers $g_i$ and $g_j$ and shrinks at both ends of this curve. This gives a nontrivial cycle $\Delta_{ij}$ that is topologically a two-sphere.}. The singularities of black objects are replaced by bubbles threaded by fluxes:
\begin{equation}\label{fluxes}
\int_{\Delta_{ij}}\Theta^I=\Pi^I_{ij}
\end{equation}
where the $\Theta^I$ are the dipole two-forms defined in~\eqref{dipole_forms} and the fluxes $\Pi_{ij}^I$ are defined in~\eqref{omega}. These fluxes prevent the various two-cycles $\Delta_{ij}$ from collapsing. Now, suppose that the number of centers $N$ and the GH charges $v_j$ subject to~\eqref{constraint1} have been fixed. Then the fluxes $\Pi_{ij}^I$ depend on the parameters $k_j^I$. Since these fluxes determine completely the distances between the colinear centers by the bubble equations~\eqref{bubbleequations2}, if one wants to increase or decrease the distances between the $N$ centers one can just modify the values of $k_j^I$. This is the key ingredient to build smooth microstates of black holes and black rings using the procedure of Appendix~\ref{appendix_background}. 

In~\cite{Bena_foaming} a smooth microstate was built for a maximally-spinning extremal horizonless BMPV black hole by taking a blob of GH centers ensuring that the $(v_i,k_i^I)$ parameters are roughly of the same order. In the same paper it was shown that if one takes a huge number of centers and randomly assigns the $(v_i,k_i^I)$ one inevitably ends up with a microstate of a maximally-spinning black hole.

On the other hand, in~\cite{Bena_bubbling_supertubes} it was shown that to build a bubbled solution for a maximally-spinning (zero-entropy) black ring it suffices to take a  blob of GH centers (at least two) with zero total $v$-charge and a far-away GH center with $v=1$. Such a solution can be obtained by choosing $v_j$ approximatively of the same order, while taking $k_j^I \sim a$ for $j=1,...N-1$ and $k_j^I <<a$ for $j=N_1$.

A crucial difference between these microstate solutions and the corresponding black holes and black rings is that the latter have an infinite $AdS$ throat, while the throat have a finite throat that ends in smooth cap. There exists furthermore a limit in which the length of the throat can become infinite, and this scaling behavior~\eqref{scaling} can be achieved for colinear solutions by tuning the $v_j$ and the $k_j^I$ \cite{Bena:mergers}, and for non-colinear solutions by changing the angles between the GH centers \cite{Bena:abyss, bates_denef}.

Scaling solutions have also proved to be necessary ingredient for building microstates for BMPV black holes with \textit{large} horizon area.  This was done in~\cite{Bena:mergers} by merging a black ring blob with a black hole blob at its center. After a suitable choice of the black hole and black ring charges, this process can results in a microstate for a BMPV black hole with large horizon area. 
 
We are now able to give a complete physical interpretation of the microstate of Section~\ref{section_review}. The parameters~\eqref{charges} are chosen so that centers $g_1$ and $g_2$ together with their counterparts $g_6$ and $g_7$ via~\eqref{z2symmetry} are far away from the central blob~\eqref{ratios}. The central blob $g_3-g_4-g_5$ has total GH charge one and hence represents a bubbled maximally-spinning black hole. The two satellites have zero GH charge, and  represent two symmetric bubbled black rings.  Because of the symmetry~\eqref{z2symmetry} the full solution has $J_1=J_2=J$, and furthermore one can check that $Q_1 Q_2 Q_2 > J^2$. Hence the solution represented in Figure~\ref{figurecolinear} can be interpreted as a microstate of a BMPV black hole with a macroscopically large horizon area.

\end{document}